\documentclass[conference,9pt]{IEEEtran}
\usepackage{cite}
\usepackage{graphicx}
\usepackage{multirow}
\usepackage{multicol}
\usepackage{color}
\usepackage{url}
\usepackage{array}
\usepackage{algorithm}
\usepackage{algpseudocode}
\usepackage{textcomp}
\usepackage{amsmath}
\usepackage{mwe}

\usepackage{amsfonts}
\usepackage[normalem]{ulem}
\usepackage{enumitem}

\definecolor{gray1}{gray}{0.7}
\definecolor{gray2}{gray}{0.98}
\definecolor{light-gray}{gray}{0.95}
\newcommand{\ignore}[1]{}

\usepackage{tikz}
\usetikzlibrary{shapes,arrows}
\usepackage{subfigure}
\usepackage{flexisym}
\usepackage{multirow}

\newcommand\blfootnote[1]{%
  \begingroup
  \renewcommand\thefootnote{}\footnote{#1}%
  \addtocounter{footnote}{-1}%
  \endgroup
}

\def\BibTeX{{\rm B\kern-.05em{\sc i\kern-.025em b}\kern-.08em
    T\kern-.1667em\lower.7ex\hbox{E}\kern-.125emX}}

\begin{document}


\title{Accelerating Electrostatics-based Global\\ Placement with Enhanced FFT Computation
}

\author{\IEEEauthorblockN{Hangyu Zhang and Sachin S. Sapatnekar}
\IEEEauthorblockA{University of Minnesota, Minneapolis, MN, USA}}

\maketitle

\begin{abstract}
Global placement is essential for high-quality and efficient circuit placement for complex modern VLSI designs. Recent advancements, such as electrostatics-based analytic placement, have improved scalability and solution quality. This work demonstrates that using an accelerated FFT technique, AccFFT, for electric field computation significantly reduces runtime. Experimental results on standard benchmarks show significant improvements when incorporated into the ePlace-MS and Pplace-MS algorithms, e.g., a 5.78$\times$ speedup in FFT computation and a 32\% total runtime improvement against ePlace-MS, with 1.0\% reduction of scaled half-perimeter wirelength after detailed placement.
\end{abstract}

\begin{IEEEkeywords}
Electrostatics-based placement, FFT acceleration, Electrical field calculation, Potential calculation, Global placement.
\end{IEEEkeywords}

\blfootnote{This work was supported in part by the National Science Foundation.}

\section{Introduction}
\label{sec:Intro}

\noindent
Placement is critical in determining power, performance, and area (PPA) metrics for an integrated circuit. The problem involves positioning cells and macros to minimize wirelength, satisfy timing, and avoid overlaps. In addition, scalability and speed are significant considerations as modern circuits contain millions of mixed-size cells.

Early placement paradigms based on stochastic optimization~\cite{TimberWolf} and partitioning~\cite{Min-cut, MP_tree} are now largely displaced by quadratic and nonlinear analytical placers. Quadratic placers~\cite{ComPLx, FastPlace, FastPlace2, FastPlace3} are fast but limited by low-order wirelength modeling. Nonlinear analytical placers (APlace3~\cite{APlace}, NTUplace3~\cite{NTUplace3}, the ePlace family~\cite{ePlace, ePlace-MS, RePlAce}, Pplace-MS~\cite{Pplace}) offer higher solution quality, using techniques such as smoothed density functions or electrostatics-based approaches. GPU-accelerated placers, such as DREAMPlace~\cite{DREAM-Place} and Xplace~\cite{Xplace}, achieve significant speed improvements by integrating machine learning (ML) infrastructure. These tools leverage ML frameworks \ignore{such as PyTorch} to parallelize computations on GPUs but do not use ML inference.

Today's advanced placers are largely based on the ePlace paradigm~\cite{ePlace,ePlace-MS,RePlAce,Pplace}, which uses electrostatic principles to induce overlap removal between cells during placement iterations. A major bottleneck in these solvers is the FFT computation that determines spreading forces. Pplace-MS~\cite{Zhu18,Pplace} proposes a novel methodology by formulating the potential energy of each block through an analytical solution to Poisson’s equation. To enhance computational efficiency, it employs a fast and approximate technique for computing the partial derivatives of the potential energy, thereby significantly reducing reliance on repeated FFT operations. 

In our work, we develop new methods to enhance the efficiency of the FFT computation step by recognizing and leveraging the difference between near-range and far-range interactions. The intuitive idea is that FFT computations are limited by the accuracy requirements of short-range interactions, but long-range interactions do not require such fine granularity. Therefore, a coarse-grained grid is first used to compute long-range interactions; then a finer grid is superposed over near regions, and the inaccurate short-range interactions on the coarse grid are replaced by accurate local calculations based on FFTs on finer subgrids.  The core idea is inspired by the precorrected FFT (PCFFT) method~\cite{PCFFT} in electrostatics. However, the PCFFT method is inefficient in computing near-range interactions for the scale of problems encountered in placement: our AccFFT method adds novel computation strategies that are effective for placement problems. Furthermore, we demonstrate that our method is compatible with GPU acceleration, enabling seamless integration into GPU-accelerated placement frameworks.  Our major contributions are:

\begin{itemize}
\item We develop an FFT computation to rapidly compute spreading forces, significantly accelerating the FFT and improving total placer runtime. We propose a novel approach that efficiently computes local FFTs for short-range interactions, overcoming a major bottleneck in prior electrostatics approaches.
\item As compared to ePlace-MS, we remove the need for potential calculation, thus reducing the computational cost,
and use a simpler method for wirelength computation.
\item We demonstrate that our FFT acceleration method is general and can also be used to accelerate Pplace-MS method.
\end{itemize}

\noindent
We validate the effectiveness of our proposed approach by comparing it with two placement frameworks: ePlace-MS~\cite{ePlace-MS}, and Pplace-MS~\cite{Pplace}. Our accelerated method, Acc-ePlace, shows $5.78\times$ speedup in FFT computation, 32\% improvement in total runtime, and 1.0\% better wirelength after detailed placement (DP) compared to the algorithm in ePlace-MS~\cite{ePlace-MS}. 
When our accelerated FFTs are incorporated into Pplace-MS, our Acc-Pplace algorithm yields $2.48 \times$ speedup in FFT computation, a 20\% improvement in total runtime with no wirelength degradation. Moreover, we demonstrate the scalability of our method by incorporating it with the GPU-accelerated method in DREAMPlace~\cite{DREAM-Place}. In this context, our method achieves $5.82 \times$ speedup in FFT computation, a 33\% improvement in total runtime, and 1\% improvement in wirelength compared to the GPU-accelerated electrostatics-based method in DREAMPlace.

The paper is organized as follows. Section~\ref{sec:Preliminary} overviews the fundamental placement problem formulation and provides an overview of our approaches for the accelerating ePlace-MS and Pplace-MS approaches. This is followed by a description of the FFT acceleration method in Sections~\ref{sec:PCFFT} and~\ref{sec:our_precorrection}. Next, Section~\ref{sec:results} presents our experimental results, and Section~\ref{sec:conclusion} concludes this paper.

\section{Placement using Electrostatic Spreading Forces}
\label{sec:Preliminary}

\subsection{Preliminaries}

\noindent
Given a circuit netlist defined by cells and nets, the objective of placement is to find positions that minimize the wirelength under density constraints. The layout area is divided into an array of rectangular bins, and the placement problem is formulated as:
\begin{equation}
\begin{aligned}
\min_{\mathbf{v}} &\ W(\mathbf{v}) \\
\text{s.t.} \quad & \rho_b(\mathbf{v}) \leq \rho_t, \quad \quad \forall b \in \boldsymbol{B}
\end{aligned}
\label{eq:minW}
\end{equation}
Here, $\mathbf{v}(x, y)$ lists the cell positions; $W(\mathbf{v})$ is the wirelength; $\rho_b(\mathbf{v})$ is the density of bin $b$ (the total cell area in $b$); $\rho_t$ is the target density. The wirelength function is best captured by the HPWL, but since this function is not differentiable, the weighted average (WA) model~\cite{WA} has been proposed as an alternative, which is given by:
\begin{align}
\text{WA}: & W_e(\mathbf{v}) = \frac{\sum_{i \in e} x_i e^{x_i/\gamma}}{\sum_{i \in e} e^{x_i/\gamma}} - \frac{\sum_{i \in e} x_i e^{-x_i/\gamma}}{\sum_{i \in e} e^{-x_i/\gamma}}
\label{eq:WA_W}
\end{align}
Here, $x_i$ is the x-coordinate of cell $i$ and $\gamma$ is the smoothing parameter that controls model accuracy. We use~\eqref{eq:WA_W} as our wirelength function.

The ePlace method models circuit placement as an electrostatic system, where each cell is treated as a positive charge with its magnitude proportional to the cell area. Placement density, represented as electric charge density, is controlled by solving Poisson’s equation, which relates electric potential, electric field, and charge distribution. The density penalty function $N(\mathbf{v})$ is defined as the potential energy,
\begin{equation}
N(\mathbf{v}) = \frac{1}{2} \sum_{i \in V} N_i(\mathbf{v}) = \frac{1}{2} \sum_{i \in V} q_i \psi_i(\mathbf{v}).
\label{eq:N(v)}
\end{equation}
Here $q_i$ and $\psi_i$ denote the electric charge (represented by the cell area) and the electric potential of charge $q_i$, respectively. Then the constrained placement problem is transformed into an unconstrained problem with the objective function:
\begin{equation}
f(\mathbf{v}) = W(\mathbf{v}) + \lambda N(\mathbf{v}),
\label{eq:f(v)}
\end{equation}
where the penalty factor $\lambda$ enforces cell density constraints and minimizes wirelength. The objective function gradient is:
\begin{equation}
\nabla{f(\mathbf{v})} = \nabla{W(\mathbf{v})} + \lambda \nabla{N(\mathbf{v})}
\label{eq:gradient_f}
\end{equation}
Here, $\nabla{W(\mathbf{v})}$ represents the wirelength gradient, and $\nabla{N(\mathbf{v})}$ denotes the gradient of the density penalty, also referred to as the electric force. Since the electric field, $\xi_i(\mathbf{v}) = \partial \psi_i(\mathbf{v}) / \partial x_i$, the gradient of density penalty with respect to the movement of the charge $i$ is~\cite{ePlace-MS}:
\begin{equation}
\nabla{N(\mathbf{v})} = q_i \xi_i(\mathbf{v})
\label{eq:gradient_N}
\end{equation}
where $\xi_i(\mathbf{v})$ denotes the electric field of charge $i$. The placer uses $N(\mathbf{v})$ to even out the cells, identifying overfilled or underfilled regions and guiding cell movement using the spreading force $\nabla{N(\mathbf{v})}$.

To solve this nonlinear optimization problem, the ePlace family~\cite{ePlace,ePlace-MS,RePlAce} employs Nesterov’s accelerated gradient method, using a dynamically estimated step length based on the inverse Lipschitz constant. This approach ensures convergence by iteratively refining the solution. \ignore{and is effective for large-scale problems.}  Pplace~\cite{Zhu18,Pplace} uses a similar electrostatic framework as ePlace, but generates spreading forces from the electric potential.

\begin{algorithm}[t]
\caption{Acc-ePlace($\mathbf{v}_0$, $\tau_{\text{min}}$, $\rho_t$, $\alpha$, $\mathbf{v_{\text{gp}}}$)}
{\small
\begin{algorithmic}[1]
\State \textbf{Input:} Initial placement $\mathbf{v}_0$, overflow threshold $\tau_{\text{min}}$, density threshold $\rho_t$, coarsening factor $\alpha$
\State \textbf{Output:} $\mathbf{v_{\text{gp}}}$
\State $\mathbf{v} = \mathbf{v}_0$
\State Initialize penalty factor $\lambda$ as in ePlace-MS~\cite{ePlace-MS} 

\While{TRUE} 
    \State Determine density distribution $\boldsymbol{\rho}$ from placement $\mathbf{v}$ \label{algo1:compute_rho}
    \State $\mathbf{q}_s$ = Coarsen($\boldsymbol{\rho}, \alpha$) \label{algo1:coarsen}
    
    \State $\boldsymbol{\xi}_s$ = AccFFT($\mathbf{q}_s$)
    \label{algo1:pcfft_end}

    \State \textit{// Optimization step} 
    \State Compute $\nabla W$ // gradient of Eq.~\eqref{eq:WA_W} \label{algo1:compute_Wgrad}
    \State $\nabla N = \mathbf{q}_s \boldsymbol{\xi}_s$ // using Eq.~\eqref{eq:gradient_N} \label{algo1:compute_Ngrad}
    \State Compute $\nabla f = \nabla W + \lambda \nabla N$ \label{algo1:compute_gradient}
    \State $\mathbf{v}$ = Nesterov($\mathbf{v}$) \label{algo1:nesterov}
    \State Update: $\lambda, \gamma$, and $\tau$, as in ePlace-MS~\cite{ePlace-MS} \label{algo1:update_parameters}
    
    \If{$\tau < \tau_{\text{min}}$} \label{algo1:check_convergence}
        \State $\mathbf{v_{\text{gp}}} = \mathbf{v}$
        \State \textbf{break}
    \EndIf
\EndWhile

\State \textbf{end}
\end{algorithmic}
}
\label{algorithm:1}
\end{algorithm}

\subsection{Overview of our approach for accelerating ePlace and Pplace}
\label{sec:overview_approach}

\noindent
Our approach for accelerating ePlace-MS is outlined in Algorithm~\ref{algorithm:1}.  The inputs to our algorithm are: an initial placement solution $\mathbf{v_0}$ (using a bound-2-bound net model~\cite{B2B_kraftwerk2}), the minimum overflow threshold $\tau_{\text{min}}$ (stopping criterion), the target density $\rho_t$, and the coarsening factor, $\alpha$. The output is the global placement solution, $\mathbf{v_{\text{gp}}}$. In each iteration, the density distribution $\boldsymbol{\rho}$ is determined (line~\ref{algo1:compute_rho}), and our accelerated FFT computation, AccFFT, is invoked (line~\ref{algo1:pcfft_end}).  The FFT for the electric field $\boldsymbol{\xi}$ computation is accelerated by using a coarsened grid to capture long-range interactions between charges (Section~\ref{sec:PCFFT}). For short-range interactions, we propose a novel approach that performs localized FFTs on a localized fine grid (Section~\ref{sec:our_precorrection}).  The gradient of the objective function is calculated by combining the wirelength gradient (line~\ref{algo1:compute_Wgrad}) with the density penalty gradient (line~\ref{algo1:compute_Ngrad}), scaled by the penalty factor $\lambda$ (line~\ref{algo1:compute_gradient})

Similar to ePlace-MS~\cite{ePlace-MS}, Nesterov’s method is used as a nonlinear solver (line~\ref{algo1:nesterov}) to update the placement. Parameters such as $\lambda$, density overflow $\tau$, and wirelength smoothing $\gamma$ are updated based on the current placement (line~\ref{algo1:update_parameters}). Iterations continue until $\tau$ falls below $\tau_{\text{min}}$ (line~\ref{algo1:check_convergence}), at which point the algorithm converges, setting $\mathbf{v_{\text{gp}}}$ to the current placement $\mathbf{v}$, and the iterative loop ends.

A significant difference from ePlace-MS is that our approach does not require the calculation of the potential. While ePlace-MS uses the computed potential to maintain boundary conditions that keep cells within the placement area, we handle boundary conditions by verifying whether the position of a cell falls outside the defined placement area. If a cell is detected to be outside the boundary, we relocate it back to the edge of the nearest grid region on the boundary using a computationally inexpensive operation. This can increase the cell density near the boundary, but the higher density then generates a repulsive force that pushes the cell further inward, encouraging a more even distribution within the placement region.

To further reduce the overall runtime, we analyzed the breakdown of runtime for global placement reported in the ePlace-MS paper~\cite{ePlace-MS} The wirelength gradient computation accounts for 29$\%$ of the total runtime, largely due to the numerous exponential calculations in Eq.~\eqref{eq:WA_W}. Although the ePlace family~\cite{ePlace, ePlace-MS, RePlAce} uses a faster exponential approximation~\cite{faster_exp}, the large number of nets still makes these calculations costly. To address this, we adopt the same approximation but further optimize it by clamping the exponential value to a maximum $e^\beta$ when $x > \beta$ during early iterations. This contributes to only $\sim$10\% of our reduction in computational overhead with minimal impact on HPWL ($\sim$$0.2\%$): the major improvement ($\sim$90\%) of our approach comes from FFT acceleration.

In summary, notable differences from ePlace include:
\begin{itemize} 
\item[(1)] the use of the accelerated techniques for FFT computation;
\item[(2)] the elimination of the potential calculation step in ePlace-MS, which enforces conditions at the placement region boundary;
\item[(3)] fast and approximate computation of the objective function in early iterations.
\end{itemize}

We also apply innovations (1) and (3) to Pplace-MS, applying accelerated FFT calculation to potential computation in Pplace, and computing the objective efficiently. As we will show in Section~\ref{sec:FFTcomp}, Pplace-MS is comparable to our approach in reducing the number of FFT computations over ePlace: Pplace computes the potential but not the field, while we compute the field (but not the potential).

\begin{table}[t]
   \centering
   \caption{Outline of the computation for ePlace-MS, Pplace-MS, and our Acc-ePlace approach. Acc-Pplace (not shown) accelerates the FFT while keeping the computational structure of Pplace-MS.}
   \vspace{-3mm}
   \label{tab:analogy}
   \hspace*{-0.025\linewidth}
   \resizebox{1.05\linewidth}{!}{
        \begin{tabular}{|l|l|l|l|}
        \hline
        & \multicolumn{1}{c|}{\bf ePlace-MS}
        & \multicolumn{1}{c|}{\bf Pplace-MS}
        & \multicolumn{1}{c|}{\bf Acc-ePlace}
        \\ \hline
        {\bf Density}      
        & \multicolumn{3}{c|}{$\rho(x,y)$ \hspace*{12mm}} 
        \\ \hline
        {\bf Coarsening}  
        & No coarsening
        & No coarsening 
        & Coarse grid: $\mathbf{q}_s$ = Coarsen($\boldsymbol{\rho}, \alpha$)    
        \\ \hline
        {\bf FFT step}     
        & \shortstack[l]{Calculating FFT 
        \\coefficients of $\rho(x,y)$ 
        \\on a fine grid}
        & \shortstack[l]{Calculating FFT 
        \\coefficients of $\rho(x,y)$ 
        \\by analytical solution}
        & \shortstack[l]{FFT for long-range interactions 
        \\ (Section~\ref{sec:PCFFT}) + correction for 
        \\
        short-range interactions (Section~\ref{sec:our_precorrection})}
        \\ \hline
        \multirow{2}{*}{\bf IFFT result} 
        & Potential $\boldsymbol{\phi}$ 
        & \multirow{2}{*}{Potential $\boldsymbol{\phi}$} 
        & \multirow{2}{*}{Field  $\boldsymbol{\xi}$} 
        \\ \cline{2-2}
        & Field  $\boldsymbol{\xi}$  
        & &                                \\ \hline
        {\bf Force on a cell} 
        & $F$ = $\boldsymbol{\xi} \times$ Area \hspace*{12mm}
        & $F$ = $-\left( \frac{\partial \mathcal{D}}{\partial x}, \frac{\partial \mathcal{D}}{\partial y} \right)$
        & $F$ = $\boldsymbol{\xi} \times$ Area \hspace*{12mm}
        \\ \hline
        {\bf Computation} 
        & 1 FFT, 3 IFFTs       
        & 1 FFT, 1 IFFT    
        & 1 FFT, 1 IFFT (both accelerated)
        \\ \hline
        \end{tabular}
    }
   \vspace{-5mm}
\end{table}

\subsection{Comparison of FFT computations with ePlace-MS and Pplace}
\label{sec:FFTcomp}

\noindent
A comparison of our approach with ePlace-MS~\cite{ePlace-MS} and Pplace-MS~\cite{Pplace}, shown in Table~\ref{tab:analogy}, illustrates the differences in how these methods manage charge density and calculate the field-critical components of the placement algorithm. As shown in the table, in all three cases, we first create a charge distribution, $\boldsymbol{\rho}(x,y)$, by representing each cell by a charge that is proportional to its area. Next, in the FFT-based technique in ePlace-MS, the frequency-domain coefficients, $a_{uv}$, of the charge density, $\boldsymbol{\rho}(x,y)$, are computed. However, in Pplace-MS, the coefficients are obtained by a truncation of the infinite series from the analytical solution shown in~\cite{Zhu18}. Based on these coefficients, the electric potential, the electric field, and the electrical force on each cell are computed in closed form based on Poisson's equation and its boundary conditions. An inverse FFT then yields the spreading force $F$ in the spatial domain, which is then used to formulate the optimization objective, $f(\mathbf{v})$, in~\eqref{eq:f(v)}. This function is optimized to yield the updated cell locations. These locations are used to update the cell densities $\boldsymbol{\rho}(x,y)$, in the next iteration, and the procedure continues until the placement is determined. In each iteration, the computed electric potential plays a role in enforcing boundary conditions and acts as a measure of how close the system is to reaching electrostatic equilibrium. As the system approaches equilibrium, the total potential energy decreases, indicating a more balanced distribution of cells. In ePlace-MS, the computed electric force guides the movement of cells in the direction of the electric field, while in Pplace-MS, the negative gradient potential
energy is treated as a force to move cells. These forces act to minimize potential energy by redistributing cells to achieve a more stable and even configuration across the placement area.

In contrast, our Acc-ePlace approach uses our accelerated AccFFT method to compute the spreading forces.  The source distribution, $\mathbf{q}_s$ is projected to a coarse grid, $\mathbf{q}_g$, whose frequency domain representation is computed using an FFT. Operating in the frequency domain, we then compute the field by multiplying the charge distribution by Green's function $H$, which depends purely on the grid structure of the layout. Next, we perform an IFFT to convert the field back into the spatial domain. The force is then calculated by multiplying the field with the charge quantity, which corresponds to the area of the cell. As stated in Section~\ref{sec:overview_approach}, our approach does not require the computation of potentials. Therefore, unlike ePlace-MS, which requires 1 FFT and 3 IFFTs, our method reduces the computational cost by requiring only 1 FFT and 1 IFFT, matching the operation count of Pplace-MS. 
Moreover, we incorporate our accelerated AccFFT method within Pplace-MS as Acc-Pplace, and demonstrate efficiency improvements by speeding up its FFTs for potential calculation.
\section{Accelerating FFT Computations using a Coarse Grid}
\label{sec:PCFFT}

\subsection{Overview}
\noindent
The ePlace and Pplace family of placers compute potential interactions using FFT on a fine-grained $M \times N$ uniform grid, achieving an $O(MN\log MN)$ complexity via convolution, instead of $O(M^2N^2)$ for pairwise computations. However, the FFT remains a major computational bottleneck.

\begin{figure}[t]
  \centering
  \includegraphics[width=0.5\textwidth]{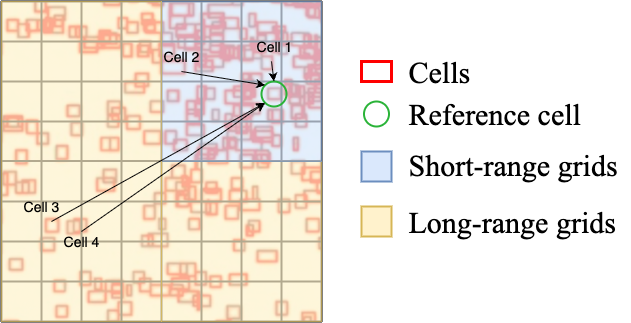}
  \caption{Long-short interactions: the red rectangular blocks represent cells. The blue grids mean that the cells in these grids have short-range interactions, while the yellow grids mean that the cells in these grids have long-range interactions with the cells in blue grids.}
  \label{fig:Long-short_interations}  
  \vspace{-6mm}
\end{figure}

Fig.~\ref{fig:Long-short_interations} shows how our approach accelerates this step by decomposing charge interactions into short-range and long-range components. The figure shows the fine grid used in ePlace or Pplace. For a specific reference cell, short-range interactions correspond to cells in blue grids (e.g., cells 1 and 2), where the $1/d^2$ term in the electric field changes rapidly, while long-range interactions in yellow grids (e.g., cells 3 and 4) are farther away, where $1/d^2$ decays slowly. For short-range interactions, a fine grid is required: using a coarse grid would cause a significant error in $1/d^2$; long-range interactions can use a coarser grid, whereby the error in approximating $1/d^2$ is small.


Under this principle, we compute long-range interactions based on the precorrected FFT (PCFFT)~\cite{PCFFT,PCFFT_Inductance_sim}, which has been used for electrostatics problems.  In typical ePlace/Pplace setups, which generally use $1024 \times 1024$ grids, the method uses a coarsened grid that can efficiently handle long-range interactions, but results in errors in computing short-range interactions.  Prior work in electrostatics~\cite{PCFFT} corrected short-range interactions using expensive pairwise calculations, which do not scale well for problem sizes seen in placement.  To overcome this, we propose a novel approach that employs localized fine-grid FFTs that maintain accuracy and computational efficiency. Our approach is aided by the fact that we calculate electric fields that decays as $1/d^2$, and therefore see a larger number of long-range interactions, as opposed to potentials in ePlace and Pplace, which decay as $1/d$.

\subsection{Computing long-range interactions}

\noindent
Our approach for computing long-range interactions follows three key steps~\cite{PCFFT}: projection, FFT computation, and interpolation.

\noindent
\textbf{Projection.}
The original charge distribution is based on a distribution of cells in the placement. The approach in ePlace/Pplace creates a fine $M \times N$ grid (typically, $M=N=1024$) and sums up all charges in each grid cell and places them in the center.  In our approach, we use a coarser $m \times n$ grid based on a \textit{coarsening ratio}, $\alpha$, so that $m=M/\sqrt{\alpha}, n = N/\sqrt{\alpha}$; typically, $M$, $N$, and $\alpha$ are chosen to be suitable powers of 2 so that $m$ and $n$ are integers.  However, simply summing up the charges within each cell and placing them in the center of the grid cell can lead to errors.  The projection step preserves accuracy based on the First Uniqueness Theorem~\cite{Griffiths13} the original charges to the coarse grid such that the projected coarse grid fields are constrained to match the fields from the original source charges at selected boundary test points known as collocation points. This condition is enforced as:
\begin{equation}
E^{gt} \mathbf{q_g}(k) = E^{st} \mathbf{q_s}(k)
\label{eq:grid_source_map}
\end{equation}
where $\mathbf{q_g}(k)$ and $\mathbf{q_s}(k)$ are, respectively the coarse grid and original charge vectors for cell $k$, and $E^{gt}$, $E^{st}$ are mapping matrices that relate coarse grid and original charges to the field at a set of chosen test points, i.e.,
\begin{equation}
E^{gt}_{ij} = \frac{\mu_0}{4\pi} \sum_{j} \frac{1}{\| \mathbf{v}^t_i - \mathbf{v}^g_j \|^2},
\; \; \;
E^{st}_{ij} = \frac{\mu_0}{4\pi} \sum_{j} \frac{\mathbf{q}^s_j}{\| \mathbf{v}^t_i - \mathbf{v}^s_j \|^2}
\label{eq:grid_source_test_map}
\end{equation}
where $\mathbf{v}^t_i$, $\mathbf{v}^g_j$, and $\mathbf{v}^s_j$ are the positions of the $i^{\rm th}$ test point, $j^{\rm th}$ coarse grid point, and $j^{\rm th}$ original source, respectively. A weight matrix $M$ distributes each original charge to the coarse grid points based on relative positions:
\begin{equation}
M(i,j) = \left[ E^{gt} \right]^\dagger E^{st,j}
\label{eq:mapping_fcn}
\end{equation}
where $\left[ E^{gt} \right]^\dagger$ denotes the pseudoinverse of $E^{gt}$, and $E^{st,j}$ refers to the $j^{\rm th}$ column of $E^{st}$. From~\eqref{eq:grid_source_map} and~\eqref{eq:mapping_fcn}, each original charge within cell $k$ is distributed across multiple grid points, providing a charge distribution on the coarse grid,
\begin{equation}
\mathbf{q_g}(k) = M \mathbf{q_s}(k)
\label{eq:grid_source_map2}
\end{equation}

\noindent
\textbf{Computing grid fields using the FFT.}
After projection to the coarse grid, the field at each grid point is computed as the convolution:
\begin{equation}
\xi_g(i,j) = \textstyle\sum_{i',j'} H(i - i', j - j') q_g(i', j').
\label{eq11}
\end{equation}
where $\xi_g(i,j)$ represents the grid field at the grid point with index $(i, j)$, while $q_g(i', j')$ denotes the grid charge located at $(i', j')$. The matrix $H$ corresponds to the Green's function (representing Coulomb's law), defined as:
\begin{equation}
H(i - i', j - j') = 
\begin{cases}
\frac{\mu_0}{4\pi \|\mathbf{v}(i, j) - \mathbf{v}(i', j')\|^2}, & \text{if } (i, j) \neq (i', j'), \\
0, & \text{otherwise}.
\end{cases}
\label{eq:Green_fcn}
\end{equation}
As in ePlace, the convolution is performed efficiently as an FFT in the frequency domain, where convolutions map onto multiplications.

\noindent
\textbf{Interpolation.}
Due to the large size of each cell in the coarse grid, the field value for each charge may be significantly different from the calculated value at the center of the grid cell. The interpolation step transfers the electric field values computed on the grid back to the original source (cell) locations, effectively reversing the projection. Leveraging the symmetry of the process, the interpolation operator is the transpose of the projection matrix $M$. The field at each cell is:
\begin{equation}
\boldsymbol{\xi}_s = M^T \boldsymbol{\xi}_g = M^T H M \mathbf{q}_s,
\label{eq:source_field}
\end{equation}
where $M$ is the weight function in~\eqref{eq:mapping_fcn}, $H$ is the Green's function in~\eqref{eq:Green_fcn}, and $\mathbf{q}_s$ is the vector of source charges.

\section{Novel Methods for Computing Short-Range Interactions}
\label{sec:our_precorrection}

\noindent
An FFT has a computational complexity of $O(MN \log(MN))$, on an $M \times N$ grid.  For placement problems with a large number of cells, the number of grid cells grows larger, leading to longer running times. Our accelerated approach in Section~\ref{sec:PCFFT} reduces the cost of the FFT computation by working with a coarser $m \times n$ grid.  

While the computations on the coarse grid are efficient and accurate for long-range interactions, they can introduce large inaccuracies for short-range interactions, e.g., within a grid cell or across neighboring grid cells. Since the number of short-range interactions is smaller than the number of long-range interactions, the method in~\cite{PCFFT,PCFFT_Inductance_sim} uses a precorrection step to overcome this inaccuracy.  This step recalculates the interactions between nearby charges by directly computing these interactions between neighboring cells $k$ and $l$ as:
\begin{equation}
\boldsymbol{\xi}_s(k, l) = M(k)^T H(k, l) M(l) \mathbf{q}_s(l)
\label{eq:source_field_fcn}
\end{equation}
The grid-based error $\boldsymbol{\xi}_g(k, l)$ is subtracted, and in~\cite{PCFFT,PCFFT_Inductance_sim}, the short-range interactions are recomputed directly as
\begin{equation}
\boldsymbol{\xi}_{\text{near}} = \textstyle \sum_{l \in \text{near}} G(k, l) \mathbf{q}_s(l)
\label{eq:near_field_fcn}
\end{equation}
Then the corrected field at the source $k$ is:
\begin{equation}
\begin{aligned}
\boldsymbol{\xi}_s(k) &= \boldsymbol{\xi}_s(k) + \boldsymbol{\xi}_{\text{near}} - \boldsymbol{\xi}_s(k, l) 
= \boldsymbol{\xi}_s(k) + \tilde{G}(k,l) \mathbf{q}_s(l)
\end{aligned}
\label{eq:pc_source_field}
\end{equation}
where
$\tilde{G}(k,l) = G(k, l) - M(k)^T H(k, l) M(l)$.

\noindent
\textbf{Quantifying the high cost of pairwise computations for short-range interactions in placement.}
For the placement problem, the number of charges involved in pairwise interactions on the coarse grid for the placement problem ($\sim$$10^6$) is much larger than the test cases ($\sim$$10^4$ charges) in this prior work in electrostatics~\cite{PCFFT,PCFFT_Inductance_sim}. This quadratic cost for pairwise recomputations is prohibitive.   

We analyze the computational cost of traditional precorrection more formally.  For each source in a given grid region, this step involves computing interactions with other sources in the same grid region and neighboring regions. If we denote the number of sources in grid region $i$ by $N_{s,i}$ and the number of interacting grids by $\nu$, then the total number of nearby sources that interact with each source is $N_{\text{near}} = N_{s,i} \cdot \nu$.  Practically, $\nu$ corresponds to the $3 \times 3$ grid regions centered about the region under consideration, and it is reasonable to assume that the number of sources in each grid region within a $3 \times 3$ window is roughly constant.  Therefore, since precorrection involves the direct computation between each source within a region and other sources in the same or nearby regions, its computational complexity is $O(N_{s,i} \cdot N_{\text{near}}) = O(N_{s_i}^2 \cdot \nu)$ per region; for a grid with $n$ regions, the complexity is $O(\sum_{i=1}^n N_{s,i}^2 \cdot \nu)$.

This computation is thus affected both by the size of the grid (a typical value is $n = 1024 \times 1024 \approx 10^6$) and by the number of charges $N_{s,i}$ within a grid region $i$. Particularly in earlier iterations, $N_{s,i}$ is quite high and imbalanced, with values often exceeding $10^3$.  The $N_{s,i}^2$ contribution to the computational complexity causes this to be a significant computational bottleneck. Finally, although $\nu$ is upper-bounded by a constant, in a $3 \times 3$ region, it increases computation by 9$\times$, i.e., nearly an order of magnitude.  We propose a more efficient method for computing short-range interactions correctly, tailored to the large problem sizes encountered during placement.  

\begin{figure}[t]
  \centering
  \subfigure[]{\includegraphics[width=0.32\linewidth]{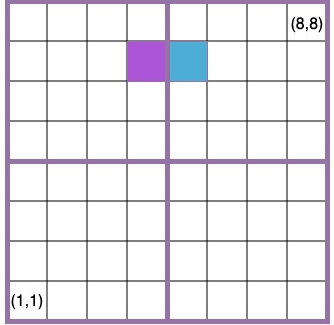}}
  \subfigure[]{\includegraphics[width=0.32\linewidth]{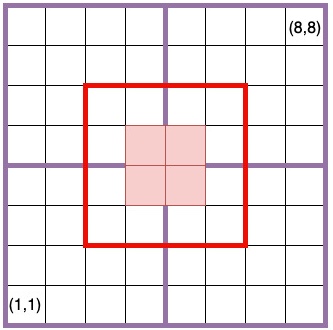}}
  \subfigure[]{\includegraphics[width=0.32\linewidth]{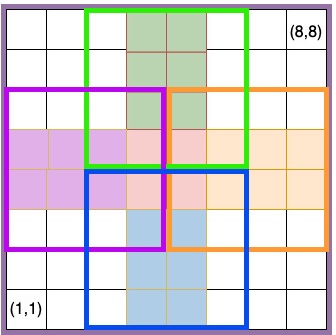}}
  \vspace{-3mm}
  \caption{An illustration of 4:1 coarsening, where one $8 \times 8$ region from the original fine grid is converted to four $4 \times 4$ regions. (a)~Four separate 16-point FFTs are used to compute short-range interactions on each of these four regions. (b)~An example showing how inaccuracies in the pink cells, which lie at the border of the four regions in (a), are resolved by computing short-range interactions within a new $4 \times 4$ region. (c)~The full set of $4 \times 4$ subregions required to cover the border regions of (a).}
  \label{fig:coarse_FFT}
  \vspace{-4mm}
\end{figure}

\noindent
{\bf Computing short-range interactions efficiently.}
We develop an alternative, computationally efficient approach for computing short-range interactions. We overcome the computational bottleneck by observing that we can achieve intermediate accuracy, between the coarse-grid FFT and full pairwise enumeration in the precorrection region by performing an FFT on the same fine grid that is applied for local precorrection.  The use of this method reduces the computational complexity to $O(r \log r)$, where $r$ is the small grid size after coarsening. This approach is more efficient and addresses the problem of increased runtime of precorrection due to the large number of sources per grid, $N_s$. We illustrate this idea in Fig.~\ref{fig:coarse_FFT}(a), which shows the original fine grid in black. The coarsened grid for long-range interactions uses a coarsening ratio of $\alpha = 4$ (line~\ref{algo1:coarsen} of Algorithm~\ref{algorithm:1}), reducing the region into four subregions in the figure. Thus, each coarse grid region (colored purple) contains $r=16$ subregions. We then find the FFT on each such $4 \times 4$ region; in the example, this corresponds to four 16-point FFTs that provide a more accurate estimate of the electric field.

This approach provides sufficient accuracy for cells in the middle of each 4×4 coarse grid region; however, edge cells are influenced by cells at the border of adjacent regions, which are not considered in the original 16-point FFT. For example, in Fig.~\ref{fig:coarse_FFT}(a), the electric field in the purple cell is affected by charges in the blue cell, but the FFT for the region with the purple cell does not include the blue cell.

Therefore, we do not use this FFT to compute interactions for cells at the border of each $4 \times 4$ region. Instead, for these cells, we perform a separate 16-point FFT that creates new $4 \times 4$ regions that place these cells at the center. Fig.~\ref{fig:coarse_FFT}(b) shows a red rectangle that encloses the four pink cells that lie at the border of the initial coarse grid. For this example, the full set of four additional $4 \times 4$ regions is shown in Fig.~\ref{fig:coarse_FFT}(c); the border cells that now lie inside each of these regions are shown by the five distinct colors in the figure. 

\ignore{
\begin{algorithm}[t]
\caption{New-precorrection}
{\small
\begin{algorithmic}[1]
\State \textbf{Input:} Original FFT
\State \textbf{Output:} FinerFFT
    \State SmallFFTs = Coarsen(FFT, $\alpha$) 
    \State Compute SmallFFTs \State FinerFFT = $\sum${SmallFFTs}
    \State Create CenterFFTs
    \State Create BorderFFTs
    \State Compute both
    \State FinerFFT = CenterFFTs $+\sum${BorderFFTs}
    \State \textbf{end}
\end{algorithmic}
}
\label{algorithm:2}
\end{algorithm}
}

\noindent
\textbf{Analyzing the cost of our fine-grained FFT computation method.}
We count the number of FFTs/IFFTs required for computing short-range interactions based on a given coarsening ratio, $\alpha$ (recall that for a coarsening ratio of $\alpha$,  the domain is divided into $\alpha$ coarse regions). This requires two types of ``small'' FFTs: (a)~at the ``four-corners'' boundaries, as in Fig.~\ref{fig:coarse_FFT}(b): the number of such FFTs/IFFTs can be shown to be $2(\sqrt{\alpha} - 1)^2$; (b)~at the horizontal and vertical boundaries, as in Fig.~\ref{fig:coarse_FFT}(c): the number of such FFTs and IFFTs can be shown to be $4\times (\sqrt{\alpha} - 1)\times \sqrt{\alpha}$. In each case, the number of regions is multiplied by 2 due to the need for both an FFT and an IFFT. Each FFT is an $(mn)$-point FFT, with $mn = MN/\alpha$ being the number of squares in each coarsened region, where $M, N, m, n$ are as defined in Section~\ref{sec:PCFFT}. The total number of FFTs/IFFTs required for computing short-range interactions is the sum, which can be simplified as $2(\sqrt{\alpha} - 1)(3 \sqrt{\alpha} - 1)$ inexpensive $(mn/\alpha)$-point FFTs/IFFTs.  For Fig.~\ref{fig:coarse_FFT}, where $\alpha = 4$, this evaluates to 10 small 16-point FFTs/IFFTs, i.e., one FFT and IFFT for each colored square in the figure, in addition to a 4-point FFT on the coarse grid. In contrast, prior approaches would perform a 64-point FFT.  

For real problems with $M = N = 1024$ grid, the benefit is even greater: prior approaches use a $10^6$-point FFT, but our AccFFT approach employs a set of localized 16-point FFTs.

\section{Experimental Results}
\label{sec:results}

\noindent
The proposed method is implemented in Python 3.10 and executed on a Linux machine with Intel(R) Xeon(R) Silver 4214 CPU @2.20GHz/256GB memory. We evaluate the method using the ISPD-2005~\cite{ISPD2005} and ISPD-2006~\cite{ISPD2006} benchmarks. To assess the placer, we compare the runtime and HPWL for different algorithms and parameter settings.

Given that our approach is built upon the foundation of electrostatic placers, we specifically targeted our comparisons against the algorithms in ePlace-MS~\cite{ePlace-MS} and Pplace-MS~\cite{Pplace} to highlight the performance gains achieved with our methods: Acc-ePlace, which accelerates ePlace-MS, and Acc-Pplace, which accelerates Pplace-MS. To fairly compare the performance of all variants reported here, all tests were executed on the same machine to ensure consistency in performance evaluation. It is important to note that variations in execution environments can sometimes lead to differences in placement solutions compared to those reported in original papers. To address this, we normalize the results to provide a clear and fair comparison between different methods.

\begin{figure}[t]
  \centering
  \subfigure[4:1]{\includegraphics[width=0.32\linewidth]{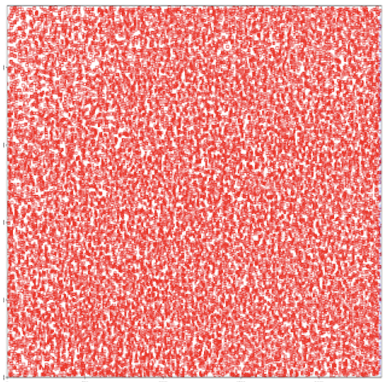}}
  \subfigure[16:1]{\includegraphics[width=0.32\linewidth]{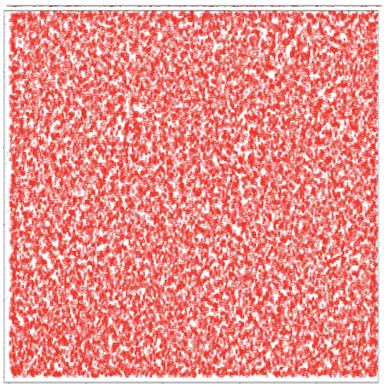}}
  \subfigure[64:1]{\includegraphics[width=0.32\linewidth]{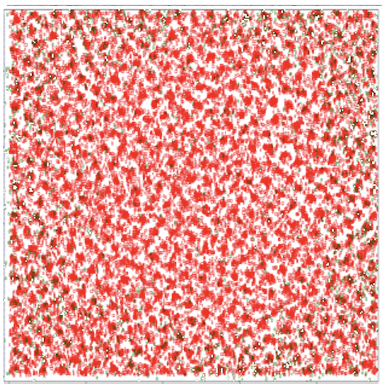}}
  \vspace{-2mm}
  \caption{The comparison of different coarsening ratios with running time and quality. (a)~With coarsening at a ratio of 4:1, the runtime achieves a speedup of $\mathbf{2.64\times}$. (b)~At a coarsening ratio of 16:1, the runtime speedup increases to $\mathbf{5.94\times}$. (c)~With coarsening at 64:1, the runtime speedup reaches $\mathbf{9.68\times}$. The cells cluster more at larger coarsening ratios.}
  \label{fig:coarse_comp}
  \vspace{-4mm}
\end{figure}

We begin by analyzing the trade-off between efficiency and accuracy associated with different coarsening ratios. Increasing the coarsening ratio yields greater runtime reduction in the computation of long-range interactions due to fewer grid points and faster computations. However, as shown in Fig.~\ref{fig:coarse_comp}, this comes at the cost of placement quality. Higher coarsening ratios lead to more tightly clustered cells and increased whitespace across the placement region, as illustrated in Fig.~\ref{fig:coarse_comp}(d), indicating that cell overlaps are not eliminated efficiently. These observations highlight the importance of selecting a coarsening ratio that balances computational efficiency and placement quality. Based on our experimental findings, we adopt a coarsening ratio of 16:1 as the optimal trade-off.

Table~\ref{tab:Comparison} compares the performance of our Acc-ePlace placement method using accelerated FFTs with the baseline placer, ePlace-MS. Columns 5--6 report results for a coarsening ratio of $\alpha=16$, showing an average speedup of $5.78\times$ in FFT computation and $1.32\times$ in total runtime. Columns 7--8 present the same comparison under GPU acceleration using the DREAMPlace framework~\cite{DREAM-Place}. Even with GPU support, our method maintains significant gains, achieving 5.82× speedup in FFT and 1.33× in total runtime. Columns 9--10 show the normalized scaled HPWL (sHPWL) results, evaluated using the official script from~\cite{ISPD2006}. Our method improves wirelength across most benchmarks, with an average improvement of $1.2\%$ after global placement (GP) and $1.0\%$ after detailed placement (DP), relative to ePlace-MS. Finally, Columns 11--14 compare our method with Pplace-MS, where our FFT acceleration is integrated into the Pplace framework. Acc-Pplace achieves a $2.48\times$ FFT speedup and a $1.20\times$ improvement in overall runtime, with comparable wirelength quality. The lower speedup relative to ePlace is attributed to the reduced number of FFTs in Pplace (Table~\ref{tab:analogy}). These results underscore the applicability of our approach across multiple placement engines.

\begin{table*}[t]
   \centering
   \vspace{-2mm}
   \caption{Comparisons between approaches that use the traditional FFT as in ePlace-MS~\cite{ePlace-MS}; GPU-accelerated method as in DREAMPlace~\cite{DREAM-Place}; Pplace-MS~\cite{Pplace}; and our accelerated methods, Acc-ePlace and Acc-Pplace, on the MMS benchmarks. The table shows the normalized FFT runtime, the normalized total placer runtime, and the normalized scaled HPWL (sHPWL) after GP and DP. Normalized sHPWL = (Acc-ePlace sHPWL)/(ePlace-MS sHPWL) or (Acc-Pplace sHPWL)/(Pplace-MS sHPWL). 
   }
   \vspace{-2mm}
   \label{tab:Comparison}
   \hspace*{-0.03\linewidth}
   \resizebox{1.05\linewidth}{!}{
      \begin{tabular}{|c||c|c|c||c|c||c|c||c|c||c|c||c|c|}
      \hline
      & \multirow{3}{*}{\textbf{\# Objects}}
      & \textbf{\#}
      & \textbf{\#}
      & \multicolumn{4}{c||}{\textbf{Speedup (vs. ePlace-MS)}}
      & \multicolumn{2}{c||}{\textbf{Normalized sHPWL}}
      & \multicolumn{2}{c||}{\textbf{Speedup (vs. Pplace-MS)}}
      & \multicolumn{2}{c|}{\textbf{Normalized sHPWL}}  
      \\  \cline{5-8}
      & 
      & \textbf{Movable}
      & \textbf{Target}
      & \multicolumn{2}{c||}{\textbf{$\alpha$ = 16 (w/o GPU)}}
      & \multicolumn{2}{c||}{\textbf{$\alpha$ = 16 (with GPU)}}
      & \multicolumn{2}{c||}{\textbf{(vs. ePlace-MS)}}
      & \multicolumn{2}{c||}{\textbf{$\alpha$ = 16}}
      & \multicolumn{2}{c|}{\textbf{(vs. Pplace-MS)}}
      \\ \cline{5-14} 
      &
      & \textbf{Macros}  & \textbf{Density}
      & \textbf{FFT}   & \textbf{Total}   & \textbf{FFT}   & \textbf{Total}   
      & \textbf{Post-GP} & \textbf{Post-DP}   
      & \textbf{FFT}   & \textbf{Total}   
      & \textbf{Post-GP}  & \textbf{Post-DP}  \\ \hline 
      ADAPTEC1  & 211,447  & 63 &  
      100$\%$ &
      5.48$\times$  &  1.37$\times$  & 
      5.55$\times$  &  1.39$\times$  &
      0.979$\times$  & 0.988$\times$  &
      2.38$\times$  &  1.18$\times$  &
      1.011$\times$      & 1.009$\times$      \\ \hline
      ADAPTEC2  & 255,023  & 127 & 
      100$\%$ &
      5.31$\times$  &  1.32$\times$  &
      5.39$\times$  &  1.34$\times$  &
      0.974$\times$  & 0.978$\times$  &      
      2.40$\times$  &  1.19$\times$  &
      0.994$\times$      & 0.996$\times$      \\ \hline
      ADAPTEC3  & 451,650  & 58 & 
      100$\%$ &
      5.81$\times$  &  1.33$\times$  &
      5.79$\times$  &  1.33$\times$  &    1.003$\times$  & 1.008$\times$  &
      2.41$\times$  &  1.23$\times$  &
      0.991$\times$      & 0.995$\times$     \\ \hline
      ADAPTEC4  & 496,054  & 69 & 
      100$\%$ &
      5.87$\times$  &   1.37$\times$  &
      5.94$\times$  &   1.38$\times$  &
      0.970$\times$  & 0.979$\times$  &
      2.44$\times$  &  1.17$\times$  &
      1.009$\times$      & 1.002$\times$     \\ \hline
      BIGBLUE1  & 278,164  & 32 & 
      100$\%$ &
      5.79$\times$  &  1.35$\times$  &
      5.81$\times$  &  1.36$\times$  &
      0.984$\times$  & 0.980$\times$  &
      2.39$\times$  &  1.16$\times$  &
      0.992$\times$      & 0.995$\times$     \\ \hline
      BIGBLUE2  & 557,866  & 959 & 
      100$\%$ &
      5.86$\times$  &  1.33$\times$  &
      5.91$\times$  &  1.35$\times$  &
      1.017$\times$  & 1.011$\times$  &
      2.47$\times$  &  1.21$\times$  &
      0.997$\times$      & 1.001$\times$     \\ \hline
      BIGBLUE3  & 1,096,812  & 2549  & 
      100$\%$ &
      5.90$\times$  & 1.32$\times$  &
      5.94$\times$  &  1.33$\times$  &
      0.978$\times$  & 0.981$\times$  &
      2.55$\times$  &  1.22$\times$  &
      1.007$\times$      & 1.004$\times$     \\ \hline
      BIGBLUE4  & 2,177,353  & 199  & 
      100$\%$ &
      5.95$\times$   & 1.28$\times$  &
      5.93$\times$  &  1.27$\times$  & 
      1.006$\times$  & 1.008$\times$  &
      2.62$\times$  &  1.20$\times$  &
      0.993$\times$      & 0.997$\times$     \\ \hline
      ADAPTEC5  & 843,128  & 76  & 
      50$\%$ &
      5.79$\times$  &  1.28$\times$  &
      5.88$\times$  &  1.30$\times$  &
      0.981$\times$  & 0.982$\times$  &
      2.54$\times$  &  1.19$\times$  &
      0.995$\times$      & 0.998$\times$     \\ \hline
      NEWBLUE1  & 330,474  & 64  & 
      80$\%$ &
      5.81$\times$  &  1.39$\times$  &
      5.84$\times$  &  1.40$\times$  &
      0.979$\times$  & 0.982$\times$  &
      2.44$\times$  &  1.20$\times$  &
      0.995$\times$      & 0.996$\times$     \\ \hline
      NEWBLUE2  & 441,516  & 3748  & 
      90$\%$ &
      5.92$\times$  & 1.35$\times$  &
      5.91$\times$  &  1.35$\times$  &
      1.008$\times$  & 1.006$\times$  &
      2.45$\times$  &  1.23$\times$  &
      1.010$\times$      & 1.007$\times$     \\ \hline
      NEWBLUE3  & 494,011  & 51  & 
      80$\%$ &
      5.69$\times$  &  1.33$\times$  &
      5.77$\times$  &  1.35$\times$  &
      0.970$\times$  & 0.976$\times$  &
      2.47$\times$  &  1.21$\times$  &
      0.992$\times$      & 0.995$\times$      \\ \hline
      NEWBLUE4  & 646,139  & 81  & 
      50$\%$ &
      5.75$\times$  &  1.30$\times$  &
      5.83$\times$  &  1.32$\times$  &
      0.971$\times$  & 0.976$\times$  &
      2.50$\times$  &  1.19$\times$  &
      0.996$\times$      & 0.997$\times$     \\ \hline
      NEWBLUE5  & 1,233,058  & 91  & 
      50$\%$ &
      5.86$\times$  & 1.28$\times$  &
      5.84$\times$  &  1.27$\times$  &
      1.005$\times$  & 1.011$\times$  &
      2.57$\times$  &  1.18$\times$  &
      1.005$\times$      & 1.002$\times$     \\ \hline
      NEWBLUE6  & 1,255,039  & 74  & 
      80$\%$ &
      5.87$\times$  & 1.25$\times$  &
      5.90$\times$  &  1.26$\times$  &
      1.010$\times$  & 1.004$\times$  &
      2.54$\times$  &  1.22$\times$  &
      0.994$\times$      & 0.995$\times$     \\ \hline
      NEWBLUE7  & 2,507,954  & 161  & 
      80$\%$ &
      5.82$\times$   & 1.19$\times$  &
      5.88$\times$  &  1.21$\times$  &
      0.974$\times$  & 0.979$\times$  &
      2.63$\times$  &  1.21$\times$  &
      0.997$\times$      & 0.998$\times$     \\ \hline \hline
      \textbf{Average}  & & & 
      & \textbf{5.78$\times$} 
      & \textbf{1.32$\times$}
      & \textbf{5.82$\times$} 
      & \textbf{1.33$\times$} 
      & \textbf{0.988$\times$} 
      & \textbf{0.990$\times$}
      & \textbf{2.48$\times$} 
      & \textbf{1.20$\times$} 
      & \textbf{0.998$\times$} 
      & \textbf{0.999$\times$} \\ \hline
      \end{tabular}
   }
   \vspace{-5mm}
\end{table*}

\begin{figure}[t]
  \centering
  \subfigure[]{\includegraphics[width=0.32\linewidth]{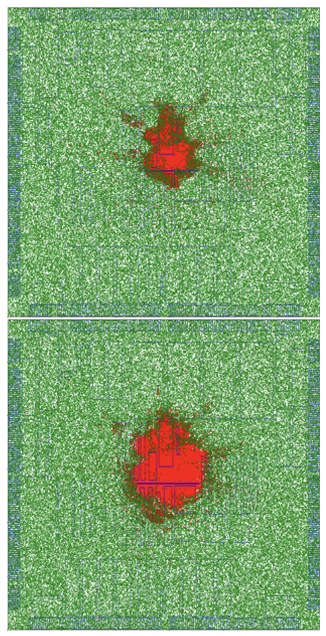}}
  \subfigure[]{\includegraphics[width=0.32\linewidth]{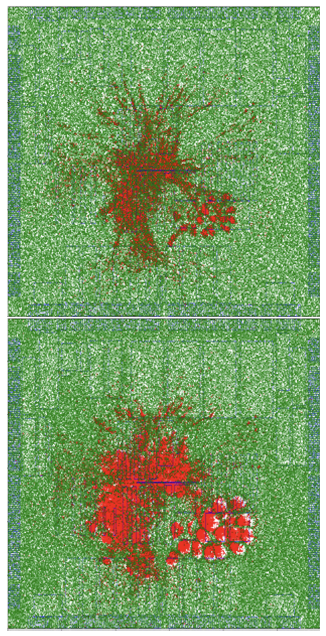}}
  \subfigure[]{\includegraphics[width=0.32\linewidth]{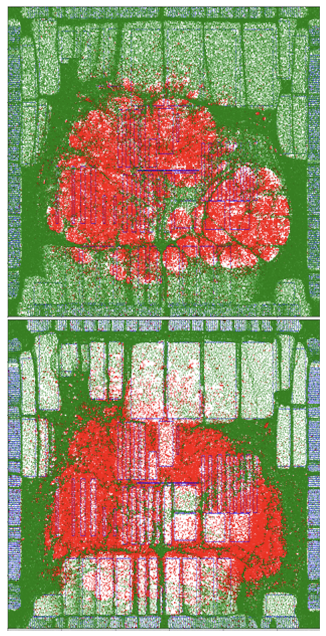}}
  \subfigure[]{\resizebox{0.45\linewidth}{!}{
  \begin{tabular}{|c|c|c|}
  \hline
  \textbf{Snapshot} & \multicolumn{2}{c|}{\# iterations}       \\ \cline{2-3}
  & ePlace 
  & Acc-ePlace  \\ \hline \hline
  (a)               & 100                & 130                 \\ \hline
  (b)               & 200                & 245                 \\ \hline
  (c)               & 400                & 500                 \\ \hline  
  \multicolumn{3}{c}{}\\
  \end{tabular}
  }}
  \vspace{-2mm}
  \caption{Iso-runtime snapshots of placement at three specific times between a placer using a traditional FFT (top) vs. our Acc-ePlace approach (bottom). Standard cells are shown as red points, macros as blue rectangles, and filler cells as green points. In the time that it takes the traditional FFT-based placer to complete (a) 100, (b) 200, and (c) 400 iterations, the Acc-ePlace placer performs more iterations, as shown in the table in (d), resulting in a larger amount of cell spreading. This leads to faster convergence of the placer, where convergence is achieved when all density constraints are satisfied.}
  \label{fig:ePlace_vs_PCFFT}
  \vspace{-6mm}
\end{figure}

Fig.~\ref{fig:ePlace_vs_PCFFT} compares the progression of placement between Acc-ePlace and the traditional FFT-based ePlace-MS for the ADAPTEC1 benchmark~\cite{ISPD2005} at iso-runtime. Standard cells, macros, and filler cells are shown as red points, blue rectangles, and green points, respectively. Our method achieves greater cell spreading within the same runtime, leading to faster convergence and reaching the target density distribution more efficiently than ePlace-MS.

\begin{table}[b]
   \centering
   \vspace{-5mm}
   \caption{Runtime analysis for conventional pairwise short-range computation vs. our FFT-based computation on localized fine grids (Section~\ref{sec:our_precorrection}) for different values of $\alpha$}
   \label{tab:direct_vs_FFT_precorrection}
    \vspace{-2mm}
    \resizebox{1\linewidth}{!}{
        \begin{tabular}{|l||cc|c||cc|c||}
        \hline
        \multirow{2}{*}{Grid size} &
        \multicolumn{2}{c|}{\begin{tabular}[c]{@{}c@{}}\% of runtime for\\ short-range interactions \\ ($\alpha = 16$)\end{tabular}} &
        \multirow{2}{*}{\begin{tabular}[c]{@{}c@{}}Total AccFFT \\ 
        runtime\\ reduction \end{tabular}} &
        \multicolumn{2}{c|}{\begin{tabular}[c]{@{}c@{}}\% of runtime for\\ short-range interactions \\ ($\alpha = 4$)\end{tabular}} &
        \multirow{2}{*}{\begin{tabular}[c]{@{}c@{}}Total AccFFT \\ runtime\\ reduction \end{tabular}} \\ \cline{2-3} \cline{5-6}
        & \multicolumn{1}{c|}{Direct} & Ours            &            & \multicolumn{1}{c|}{Direct} & Ours &      \\ \hline
        $512 \times 512$   & \multicolumn{1}{c|}{78.6\%}        & 33.2\%     & 68.5\% & \multicolumn{1}{c|}{71.2\%}        & 36.6\%     & 54.6\% \\ \hline
        $1024 \times 1024$ & \multicolumn{1}{c|}{75.5\%}        & 31.6\%     & 64.3\% & \multicolumn{1}{c|}{69.6\%}        & 34.7\%     & 53.5\% \\ \hline
        $2048 \times 2048$ & \multicolumn{1}{c|}{73.9\%}        & 30.1\%     & 63.1\% & \multicolumn{1}{c|}{68.4\%}        & 33.3\%     & 52.6\% \\ \hline
        \end{tabular}
    }
\end{table}

Furthermore, we evaluate the efficiency of our Acc-ePlace approach compared to conventional electrostatics methods for the computation of short-range interactions. Our method replaces direct pairwise interaction computations with a sequence of smaller, efficient FFTs (as detailed in Section~\ref{sec:our_precorrection}). Table~\ref{tab:direct_vs_FFT_precorrection} presents runtime comparisons across different grid sizes for coarsening ratios of 4:1 and 16:1. We report the runtime for computing short-range interactions as a percentage of total FFT evaluation time, and the overall runtime reduction for the FFT computation. The table demonstrates significant efficiency gains.

Across all grid sizes, the direct approach consistently consumes a larger proportion of Acc-ePlace runtime compared to our FFT-based method. For 4:1 coarsening, the direct method uses $68.4\%-71.2\%$ of runtime, whereas our approach reduces this to $33.3\%-36.6\%$. Across all grid sizes, 16:1 coarsening achieves greater runtime reductions than 4:1. However, based on our tests, higher coarsening ratios negatively affect placement quality by increasing cell clustering and white space within the placement area. These findings highlight the need for a balanced approach when selecting the coarsening ratio to minimize runtime while maintaining placement quality.
\section{Conclusion}
\label{sec:conclusion}

\noindent
This article has introduced a global placement algorithm based on accelerating FFT computation by separating long-range and short-range interactions. Our experimental results on the benchmarks demonstrate that the proposed algorithm outperforms existing state-of-the-art electrostatics-based mixed-size placement techniques in terms of runtime, at similar quality. The proposed method is broadly applicable to electrostatics-based placers that rely on FFTs for computing potential and electric fields, such as ePlace-MS~\cite{ePlace-MS} and Pplace-MS~\cite{Pplace}. Furthermore, it is fully compatible with GPU-accelerated frameworks like DREAMPlace~\cite{DREAM-Place}, enabling enhanced performance across both CPU-based and GPU-based placement engines.

\clearpage
\bibliographystyle{./misc/IEEEtran}
\bibliography{./bib/main}

\end{document}